\documentclass[
]{aa}

\usepackage{graphicx,natbib}
\usepackage{txfonts}
\usepackage{dcolumn}
\usepackage{lscape}
\usepackage{amsmath}
\usepackage{booktabs}
\usepackage{color, colortbl}
\definecolor{ratgray}{gray}{0.9}
\definecolor{dblue}{rgb}{0.0, 0.0, 0.65}
\usepackage{multirow}
\usepackage[
    pdfmenubar=true,
    colorlinks=true,
    linkcolor=dblue,
    urlcolor=dblue,
    citecolor=dblue
]{hyperref}

\newcommand{\um}{\,\mu\textrm{m}}

\begin{document}

   \title{A near-infrared interferometric survey of debris-disc stars.}

   \subtitle{V. PIONIER search for variability.\thanks{Based on observations made with ESO Telescopes
   at the La Silla Paranal Observatory under program IDs 088.C-0266, 089.C-0365, 090.C-0526, 091.C-0576, 091.C-0597,
   094.C-0232, and commissioning data.}}

   \author{S. Ertel \inst{1,2}
          \and
          D. Defr\`ere \inst{2,3}
          \and
          O. Absil\inst{3}\fnmsep\thanks{F.R.S.-FNRS Research Associate}
          \and
          J.-B. Le Bouquin \inst{4,5}
          \and
          J.-C. Augereau \inst{4,5}
          \and
          J.-P. Berger \inst{6}
          \and
          N. Blind \inst{7}
          \and
          A. Bonsor \inst{8}
          \and
          A.-M. Lagrange \inst{4,5}
          \and
          J. Lebreton \inst{9,10}
          \and
          L. Marion \inst{3}
          \and
          J. Milli \inst{1}
          \and
          J. Olofsson \inst{11}
          }

   \institute{
              European Southern Observatory, Alonso de Cordova 3107, Vitacura, Casilla 19001, Santiago 19, Chile
         \and
              Steward Observatory, Department of Astronomy, University of Arizona, 993 N. Cherry Ave, Tucson, AZ 85721,
              USA\\
              \email{sertel@email.arizona.edu}
         \and
              Space sciences, Technologies and Astrophysics Research (STAR) Institute, Universit\'e de Li\`ege,
              19c All\'ee du Six Ao\^ut, B-4000 Li\`ege, Belgium
         \and
              Univ. Grenoble Alpes, IPAG, F-38000 Grenoble, France
         \and
              CNRS, IPAG, F-38000 Grenoble, France
         \and
              European Southern Observatory, Karl-Schwarzschild-Stra{\ss}e 2, 85748 Garching, Germany
         \and
              Observatoire de Gen\`eve, Universit\'e de Gen\`eve, 51 ch. des Maillettes, CH-1290 Versoix, Switzerland
         \and
              Institute of Astronomy, University of Cambridge, Madingley Road, CB3 0HA, UK
         \and
              Infrared Processing and Analysis Center, California Institute of Technology, Pasadena, CA 91125, USA
         \and
              NASA Exoplanet Science Institute, California Institute of Technology, 770 S. Wilson Ave., Pasadena, CA
              91125, USA
         \and
              Instituto de F\'isica y Astronom\'ia, Facultad de Ciencias, Universidad de Valpara\'iso, Av. Gran
              Breta\~na 1111, Playa Ancha, Valpara\'iso, Chile
              }



  \abstract
   {Extended circumstellar emission has been detected within a few 100\,milli-arcsec around $\gtrsim$10\% of nearby
   main sequence stars using near-infrared interferometry. Follow-up observations using other techniques, should they
   yield similar results or non-detections, can provide strong constraints on the origin of the emission. They can
   also reveal the variability of the phenomenon.}
   {We aim to demonstrate the persistence of the phenomenon over time scales of a few years and to search for variability
   of our previously detected excesses.}
   {Using VLTI/PIONIER in $H$~band we have carried out multi-epoch observations of the stars for which a near-infrared
   excess was previously detected with the same observing technique and instrument. The detection rates and distribution
   of the excesses from our original survey and the follow-up observations are compared statistically. A search for
   variability of the excesses in our time series is carried out based on the level of the broadband excesses.}
   {In 12 of 16 follow-up observations, an excess is re-detected with a significance of $>2\sigma$, and in 7 of 16
   follow-up observations significant excess ($>3\sigma)$ has been re-detected. We statistically demonstrate with very
   high confidence that the phenomenon persists for the majority of the systems. We also present the first detection
   of potential variability in two sources.}
   {We conclude that the phenomenon responsible for the excesses persists over time scales of a few years for the
   majority of the systems. However, we also find that variability intrinsic to a target can cause it to have
   no significant excess at the time of a specific observation.}


   \maketitle
%

\section{Introduction}

The detection of circumstellar near-infrared (near-IR) excess emission at the level of $\sim$1\% within a few
100\,milli-arcsec (mas) around nearby, mature main-sequence stars, remains enigmatic. It is generally attributed
to the presence of hot, circumstellar dust. The detections have been made using near-IR interferometry
mostly employing the instruments FLUOR (Fiber Linked Unit for Optical Recombination) at the CHARA array (Center for
High Angular Resolution Astronomy; e.g., \citealt{abs06, abs13}) and PIONIER (Precision Integrated Optics Near
Infrared ExpeRiment) at the VLTI (Very Large Telescope Interferometer; \citealt{def12, ert14b}). These very
accurate instruments are pushed by these observations to their limits in terms of both statistical accuracy and ability
to calibrate the data obtained. Until now, only two detections could be confirmed from repeated observations: Vega
\citep{abs06, def11} and $\beta$\,Pic \citep{def12, ert14b}.

Mid-infrared (mid-IR) nulling observations reveal no correlation between near-IR and mid-IR excesses \citep{men14} and
follow-up observations of the near-IR excess stars, attempting to detect polarized scattered light emission from the
circumstellar dust did not result in significant detections \citep{mar16}. When combined with the near-IR detections,
these data provide strong and valuable constraints -- even in case of upper limits -- on the emission at different
wavelengths and spatial scales, and thus on the origin of the excesses \citep{leb13}. However, variability of the
excesses needs to be characterized or ruled out. In case of non-detections in follow-up observations, the original
detections need to be confirmed and it needs to be established that the excesses persist from the original detections
to the follow-up observations. At the same time,
the detection and analysis of variability can inform us regarding the origin of the emission: Theoretical models face
severe problems to explain the large amounts of dust in the innermost regions of these systems, needed to produce the
excess \citep{bon12, bon13b, bon14}. The short orbital period and high surface density are thought to result in rapid
removal of the dust from the systems by the stellar radiation pressure \citep{bac93}. The detection or not of variability
can enable us to distinguish between continuous, episodic, and catastrophic dust production.

In this paper, we present new data obtained with VLTI/PIONIER. We re-observed several times the stars for which an
excess was previously detected with this instrument and observing technique \citep{ert14b}, with the goal to demonstrate
the persistence of the phenomenon over time and to search for variability. We summarize our observing strategy and
data processing in Sect.~\ref{sect_data}. In Sect.~\ref{sect_res} we present our analysis of the detections and non
detections. We statistically show that the detection rate for our follow-up observations of known excess stars is
significantly higher than for our original survey of stars without previous information on the presence or absence of
near-IR excess. In Sect~\ref{sect_variability} we present a search for variability in the broadband excesses of single
objects and report on the first detection of potential variability in two of our targets.
In Sect~\ref{sect_remaining} we discuss possible statistical and systematic effects and argue that they are very
unlikely to produce this result. We present our conclusions in Sect.~\ref{sect_summary}.

\section{Data acquisition and processing}
\label{sect_data}

\subsection{Observations}
\label{sect_observations}

We re-observed in $H$~band several times six of the nine stars with nominal detections and one of the three stars
with tentative detections from our original PIONIER survey \citep{ert14b}. We focus here on this clean sample
of stars with excesses detected using the same instrument and technique and in the same band ($H$~band) we employ for
our follow-up
observations. Only for these targets we can confidently expect a re-detection and directly compare our detection
statistics from our original PIONIER survey with our follow-up observations. The new data presented in this work were
obtained in August~2013 and October~2014. In addition, we consider data of HD\,172555 obtained in April~2014 in the context
of a dedicated study. We compare our detection rate and excess levels from the new observations with our original
survey \citep{ert14b} and previous observations of $\beta$\,Pic (HD\,39060, \citealt{def12}). The targets and the observing dates
are listed in Table~\ref{tab_log_data}.

\begin{table}
\caption{Observing log, excesses, and variability.}
\label{tab_log_data}
\centering
\begin{tabular*}{\linewidth}{l@{\extracolsep{\fill}}lcc}
\toprule
HD     & Night & $f_\textrm{CSE}$ [\%] & $\Delta_\textrm{CSE}$ \\
\midrule
2262   & 2012-10-15$^{(1)}$ & $0.67\pm0.17\pm0.06$ & 0.91  \\
       & 2013-08-10         & $0.40\pm0.15\pm0.06$  & -0.45 \\
       & 2014-10-11         & $0.42\pm0.15\pm0.06$  & -0.34 \\
\midrule
7788   & 2012-07-23$^{(1)}$ &  $1.43\pm0.16\pm0.05$ & 3.0   \\
       & 2013-08-10         &  $0.07\pm0.15\pm0.05$ & -4.12 \\
       & 2014-10-11         &  $1.16\pm0.17\pm0.05$ & 1.55  \\
\midrule
20794  & 2012-12-16$^{(1)}$ &  $1.64\pm0.26\pm0.26$ & 1.59  \\
       & 2013-08-10         &  $0.75\pm0.20\pm0.24$ & -0.65 \\
       & 2014-10-12         &  $0.77\pm0.24\pm0.18$ & -0.60 \\
\midrule
28355  & 2012-12-15$^{(1)}$ & $0.88\pm0.08\pm0.05$  & 1.0   \\
       & 2014-10-11         & $0.52\pm0.12\pm0.05$  & -1.63 \\
\midrule
39060  & 2010-12-04$^{(2)}$ & \multirow{2}{*}{$1.48\pm0.20\pm0.05$} & \multirow{2}{*}{0.45} \\
($\beta$\,Pic) & 2010-12-20$^{(2)}$ &                &       \\
       & 2011-11-02         & $1.32\pm0.15\pm0.05$  & -0.32 \\
       & 2012-10-16$^{(1)}$ & $0.88\pm0.22\pm0.05$  & -2.05 \\
       & 2013-08-10         & $1.81\pm0.38\pm0.05$  & 1.11  \\
       & 2014-10-11         & $1.47\pm0.11\pm0.05$  & 0.64  \\
\midrule
172555 & 2012-07-24$^{(1)}$ & $0.55\pm0.25\pm0.05$  & -0.59 \\
       & 2013-04-18         & $0.93\pm0.59\pm0.05$  & 0.34  \\
       & 2014-10-12         & $0.80\pm0.18\pm0.05$  & 0.32  \\
\midrule
210302 & 2012-07-24$^{(1)}$ &  $0.83\pm0.24\pm0.05$ & 2.77  \\
       & 2013-08-10         & $-0.16\pm0.13\pm0.05$ & -1.48 \\
       & 2014-10-11         &  $0.11\pm0.14\pm0.05$ & 0.11  \\
\bottomrule
\end{tabular*}
\tablefoot{$^{(1)}$ Survey detection \citep{ert14b}. $^{(2)}$ Date obtained on 2010-12-04 and on 2010-12-20
are combined to one measurement in order to improve the accuracy since no variability is seen between the two observations.\\
$f_\textrm{CSE}$ is the flux ratio between the circumstellar emission and the star. $\Delta_\textrm{CSE}$ is the
significance of the deviation of this value from the error weighted mean of all considered measurements of this target
(Eq.~\ref{eq_delta_f_cse}). Uncertainties on the flux ratios are separated in statistical errors (first value) and
systematic errors (see Sect.~\ref{sect_remaining} for details). The two values should be added in quadrature to obtain
the total uncertainties used in this work.}
\end{table}

For our observations, we followed closely the strategy motivated and outlined in \citet{ert14b}, which we only briefly
summarize here. All observations were carried out in $H$~band using the PIONIER beam combiner on the VLTI in combination
with the 1.8\,m Auxiliary Telescopes in the compact configuration (baselines between 11\,m and 36\,m). We simultaneously
obtained squared visibility measurements on six baselines and closure phase measurements on four telescope
triplets with each science observation. A sequence of three observations on a science target was taken, bracketed and
interleaved by observations of calibrators (CAL-SCI-CAL-SCI-CAL-SCI-CAL). At least three different calibrators were
selected for each sequence and each science observation was bracketed by two different calibrators. The calibrators
were selected from the catalog of \citet{mer05}. Observations were carried out in SMALL spectral resolution (three
channels across the $H$~band). The FOWLER read-out mode and the fast AC mode were used and the number of steps
read in one scan (NREAD) was set to 1024 with a scan length of $60\um$.

The observing conditions were in general well suited for our observations (seeing and coherence time $<1.5''$
and $>2$\,ms, respectively, thin clouds at most). Only in one night, \mbox{9-Aug-2013}, the conditions were highly
variable with occasionally very large seeing values ($>2''$) and short coherence time ($\sim$1\,ms). A detailed
discussion of the systematic effects produced by such observing conditions is presented in the
Appendix~\ref{app_bad_obs}. We discard all data taken during this night from our statistical analysis, because
they have to be considered unreliable as discussed in the appendix.

\begin{figure*}
 \centering
 \includegraphics[angle=0,width=\linewidth]{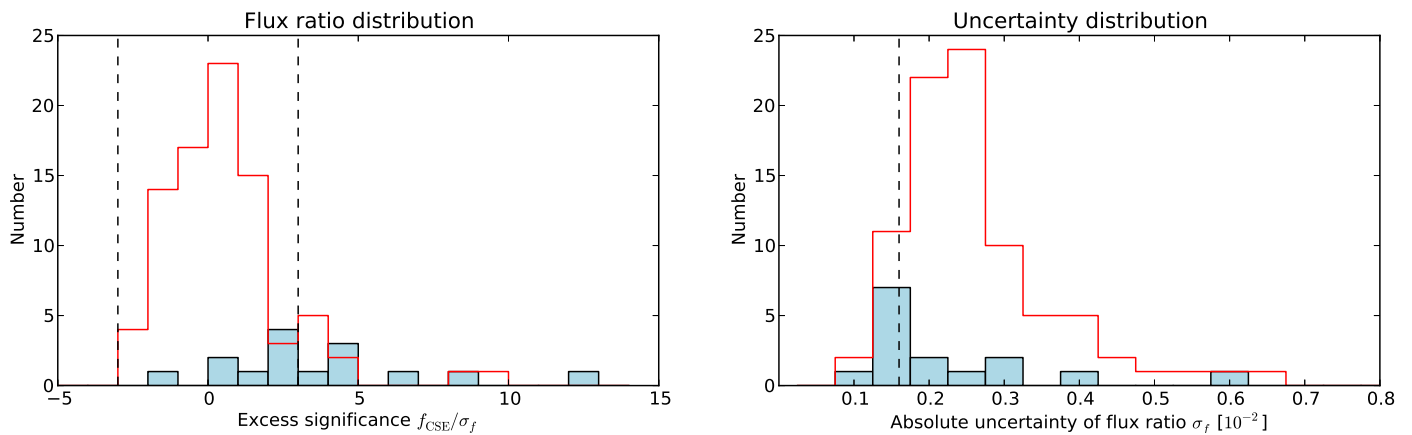}
 \caption{Excess distribution (\emph{left}) and distribution of uncertainties on the disk-to-star flux ratio (\emph{right}).
 The blue histogram represents our $H$~band follow-up observations of previous $H$~band detections
 with PIONIER. The red line shows the distribution for our original survey \citep{ert14b}. Vertical dashed lines are
 plotted at $f_\textrm{CSE}/\sigma_f = -3$ and $f_\textrm{CSE}/\sigma_f = +3$ for the excess distribution and at the
 median uncertainty ($1.6\times10^{-3}$) of our follow-up observations for the uncertainty distribution.}
 \label{fig_hist_excess}
\end{figure*}

\subsection{Data reduction, calibration, and excess measurement}
\label{sect_processing}

We reduce our new data using the standard PIONIER pipeline version~3.30 \citep{lebou11}. As for our observing strategy,
for the calibration and excess measurements we followed closely the procedure motivated and outlined by \citet{ert14b}.
First, a global calibration of each night was performed to correct for the effect of field rotation on the instrumental
visibilities (due to polarization effects in the VLTI optical train; \citealt{lebou12, ert14b}). Then, we selected
pairs of a science observation and the preceding or following calibrator observation without using the same calibrator
observation to calibrate two science observations. We also avoid using different observations of the same calibrator
for the calibration of different observations of one science target. Calibrator observations with large noise,
systematically low squared visibilities in most baselines compared to the other calibrators (indicative of
a companion or extended circumstellar emission that might be caused, e.g., by stellar mass loss on the post-main
sequence) or with signs of closure phase signal (indicative of a companion) were excluded. Both renders a
calibrator unusable.
Finally, we fit our simple exozodiacal dust model (a homogeneous emission filling the entire field of view) to all
squared visibility data obtained in one observing sequence of a target to measure the flux ratio $f_\textrm{CSE}$
between circumstellar emission and star (disk-to-star flux ratio) and its uncertainty $\sigma_f$. See \citet{ert14b}
for details about the procedure and for the stellar photometry and parameters used.

The survey data have originally been reduced using the PIONIER pipeline version~2.51. We re-reduced and calibrated
these data using the pipeline version 3.30 and found consistent results. In order not to have different (but fully
consistent) numbers in the literature for these observations, we use here the results from \citet{ert14b}. The data
obtained for $\beta$\,Pic before 2012 have been published by \citet{def12}. These observations do not follow our
optimized observing strategy, but significantly more data have been taken during each run (typically half a night
dedicated to one target). The calibration and analysis performed in \citet{def12} has been optimized for these data.
Since we do not see any reason to update these procedures, and in order not to have different (but fully consistent)
numbers in the literature, we rather re-use these results.

\section{Results}
\label{sect_res}

\subsection{Persistence of the excesses}
\label{sect_redetection}

In order to demonstrate the persistence of our detections, we would ideally like to re-detect every
excess in each observation. However, this is unrealistic, since most of our detections are close to our sensitivity
limits. An excess originally detected close to the detection threshold may be measured to be below this threshold in
a second observation with the same sensitivity due to statistical errors. Thus, a non-detection of significant excess
does not necessarily imply that the excess is no longer present. In fact, we only re-detect the excesses in $\sim$50\%
of our follow-up observations at a significance $>3\sigma$. In our original survey of 92 stars (out of which 85 were
used to derive clean statistics), we found an excess detection rate of $10.6^{+4.3}_{-2.5}$\%. We find a mean of the
excess significance $\chi_\textrm{CSE} = f_\textrm{CSE}/\sigma_f$ of $\bar{\chi}_\textrm{CSE} = 0.54$ from our
original survey and of $\bar{\chi}_\textrm{CSE} = 3.70$ from our follow-up observations.

To test if the difference in $\chi_\textrm{CSE}$ between the tow samples is statistically significant, we use a two
sample Anderson-Darling (AD) test \citep{schol87} to rule out that the distribution of $\chi_\textrm{CSE}$ from our
original survey and that from our
follow-up observations (including multiple observations of a target) are statistically consistent. If they were found
to be consistent, this would be an indication that the differences are simply caused by statistical fluctuations in
our data. Furthermore, this would be an indication that our detections are caused by imperfectly understood statistical
errors  that are not repeatable for a given observation but cause false detections with the same probability in repeated
observations. If $\chi_\textrm{CSE}$ is found to be significantly higher among the stars observed during our follow-up
campaigns, this would mean that an excess is indeed present and persistent over time for at least the majority of our
detections.

For almost all targets the original detection was made during our original survey. Only for $\beta$\,Pic the first
detection was made during two nights in December~2010 and one night in November~2011. \citet{def12} combined all these data
to measure the excess with the best accuracy (no significant variability was found), but the excess was nominally
detected in all data sets. Here, we consider the detection in the two nights in December~2010 as the original detection
and each later observation (including the observations in November~2011 and from our original survey) as follow-up
observations.

The distribution of $\chi_\textrm{CSE}$ for our original survey and our follow-up observations, are shown
in Fig~\ref{fig_hist_excess} (left panel). The AD test yields a probability of only $5.7\times10^{-5}$ that these two
samples are drawn from the same distribution, which allows us to reject this hypothesis. The right panel of
Fig.~\ref{fig_hist_excess} shows the distribution of $\sigma_f$, illustrating that the sensitivity of our follow-up
observations (mean $\bar{\sigma}_f = 0.22\%$, median $\tilde{\sigma}_f = 0.16\%$) is similar to that of our original
survey ($\bar{\sigma}_f = 0.26\%$, $\tilde{\sigma}_f = 0.24\%$).

The excess around $\beta$\,Pic is our clearest detection but has been hypothesized to originate from forward scattering
in the outer, edge-on seen disk \citep{def12}. To test the impact of this potential false positive, we repeat the AD
test excluding this star and still find a probability of only $2.2\times10^{-3}$ that the two samples are drawn from
the same distribution. $\beta$\,Pic with its massive,
young, edge-on seen debris disk is the only plausible candidate for such a false detection and even for this star
\citet{def12} rule out that more than 50\% of the excess can be produced by forward scattering in the outer disk. We
thus reject with very high confidence our null hypothesis that the distributions of $\chi_\textrm{CSE}$ from our survey
and follow-up observations are drawn from the same distribution. We thus conclude that an excess was still present and
persistent around the majority of our targets during the follow-up observations.

\subsection{Variability in single targets}
\label{sect_variability}

We have demonstrated that for a significant fraction of our targets the excess persists over time scales of a few years.
However, our analysis does not allow us to characterize or rule out variability of single sources. In the following,
we present a search for variability of the detected excesses. We focus on the broadband excesses (integrated over the
three spectral channels), where variability
is most readily detectable due to the higher significance of the detections compared to the spectrally dispersed data.
A more sophisticated search for variability including the spectral slope of the emission requires detailed modeling
of the systems and depends on model assumptions. We defer this analysis together with the production of sensitive upper
limits on targets without detected variability and a theoretical interpretation of the results to a forthcoming,
dedicated paper.

\begin{figure}
 \centering
 \includegraphics[angle=0,width=\linewidth]{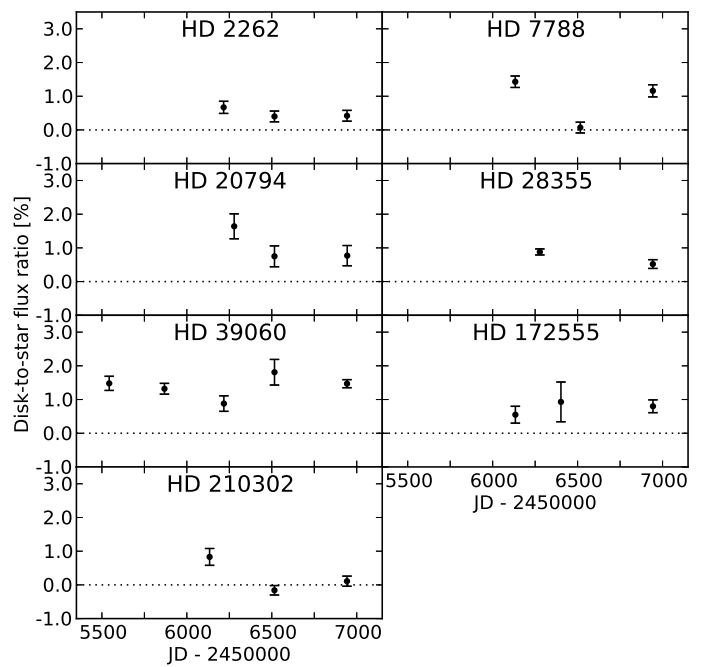}
 \caption{Time series of the excesses.}
 \label{fig_variability}
\end{figure}

Our time series of the excess measurements are plotted in Fig.~\ref{fig_variability}. We check whether the
single excess measurements for a target deviate significantly from their error weighted mean. The significance
$\Delta_\textrm{CSE}$ of this deviation is computed as
\begin{equation}
 \label{eq_delta_f_cse}
 \Delta_{\textrm{CSE},i} = \frac{f_{\textrm{CSE},i} - \left<f_\textrm{CSE}\right>}{\sqrt{\sigma_{f,i}^2 + \sigma_{\left<f\right>}^2}}
\end{equation}
where $f_{\textrm{CSE},i}$ and $\sigma_{f,i}$ are the flux ratio from a single measurement and its error, $\left<f_\textrm{CSE}\right>$
is the error weighted mean of all measurements of one target, and $\sigma_{\left<f\right>}^2$ is the standard deviation
of this mean. We again discard from our analysis the data excluded in Sect.~\ref{sect_redetection}. A significant
deviation ($> 3\sigma$) is found for one target, HD\,7788 (Table~\ref{tab_log_data}).

We emphasize that this is a simple but conservative metric. It requires, however, that the all errors are well
understood (see discussion in Sect~\ref{sect_remaining}). A statistical test of the distribution of the excess measurements
for a given target against a normal distribution would be a more sensitive tracer of variability, but is not
yet possible due to the limited number of points available for each target. We note that for HD\,210302 the measurements
from the nights of 24-Jul-2012 and 10-Aug-2013 deviate from each other by 3.5~times their respective error bars
added in quadrature. The first measurement shows significant excess ($3.3\sigma$) while the latter one and the one
obtained on 11-Oct-2014 are consistent with no excess. This may thus be considered as a tentative indication that the
excess has fainted below our sensitivity between July~2012 and August~2013. However, the largest $\Delta_\textrm{CSE}$
we find for this star is only 2.77 and we thus consider this variation not significant. For all other targets, the broadband excess
measurements are consistent with constant excess over the period it was monitored.

We conclude that with HD\,7788 we found the first strong candidate for significant variability of the faint
near-infrared excess around a nearby main sequence star. As can be seen in Fig.~\ref{fig_variability}, the excess
disappears (given our sensitivity) from the first detection to the second observation about one year later and is
re-detected about another year later.

\section{Discussion}
\label{sect_remaining}

The results from the AD test show that the detection rate from our follow-up observations of stars with
previously detected excesses is significantly higher than from our original survey of stars without previous information
on the presence of near-IR excesses. We concluded in Sect.~\ref{sect_redetection} that this is evidence for a persistence
of the excesses around the majority of our targets over time scales of several months to a few years. This conclusion
based on the statistics from our whole samples is only valid if repeated false detections around specific
targets can be ruled out as a cause for the higher detection rate. If the statistical errors are well understood,
they will not produce a significant number of false detections. If they were underestimated, they had been expected
to produce false detections with the same probability in any of our observation. They had not been reproducible for
a given target during different observing nights. Thus, effects that fall into this category, such as an underestimation
of the piston noise in our data, can be ruled out as a cause of the higher detection rate in our follow-up
observations based on the AD test results. Other errors that are not reproducible from one observation to another
(at a random night, time of the night, and observing condition), such as the presence of an unknown effect of seeing,
coherence time, or pointing direction of the observations, or the quality of the alignment of the instrument for the
observing night can be ruled out based on the same arguments.

It remains to be excluded any systematic effects common to most observations of our excess targets. Such effects
could be related to the science target itself or the calibrators used
(underestimation or overestimation of the stellar diameter of the science target or calibrator, respectively).
Systematics related to specific calibrators can result in repeatable errors, since the same calibrators
(the best ones available) were used for most of the observations of a given science target. Calibrators with faint
companions or extended circumstellar emission (bad calibrators) would however only reduce the detected excesses, not
cause false detections.

The targets for which excess has been detected and associated calibrators are not particularly bright or have large
diameters compared to our whole survey sample. Global systematics in estimating the diameters of both science targets
and calibrators would result in a global shift in the distribution of $\chi_\textrm{CSE}$ toward positive or negative
excesses, which is not seen in our survey statistics. Furthermore, uncertainties on stellar diameters are minimized by
observing at short baselines, where both science targets and calibrators are marginally resolved at most and the remaining
uncertainties are well considered in our excess and error estimation. In Table~\ref{tab_log_data}, we list
the flux ratio derived from all data taken in one observing sequence on a target together with its uncertainty
separated in statistical and systematic errors. The flux ratio $f_\textrm{CSE}$ is related to the ratio
$V^2_\textrm{meas}/V^2_\textrm{pred}$ between measured and the predicted predicted squared visibility following
\begin{equation}
 f_\textrm{CSE} \approx \frac{1}{2} \left(1 - \frac{V^2_\textrm{meas}}{V^2_\textrm{pred}}\right)
\end{equation}
\citep{dif07}. Statistical errors are estimated
from the scatter of the single measurements in one observing sequence on a science target using bootstrapping
\citep{def12, ert14b}. They represent the combined uncertainties on measuring the raw visibilities due to piston and
photon noise and due to apparent noise in the transfer function attributed to the potential presence of bad calibrators
in our data (M\'erand et al., in prep.; Ertel et al., in prep.). The systematic uncertainties represent the contribution
from uncertain diameters of our science targets and calibrators and a minor contribution from the chromaticism of
the instrument \citep{ert14b}. For HD\,20794 the uncertainties on the photometry used to estimate the stellar diameter
from surface brightness relations result in a large uncertainty on the stellar diameter. For all other stars we find
the statistical uncertainties to dominate the systematic ones. We thus consider any of the discussed effects to be
very unlikely a cause for false detections in our data.

Above, we have argued that statistical or systematic errors are very unlikely to explain the repeated detection
of excess around our science targets. This discussion, however, does not rule out that systematic errors related, e.g.,
to pointing direction, elevation, or instrument alignment (observing night) can produce a spurious variability of the
signal detected. This could lead to a false detection of variability in the case of HD\,7788. In \citet{ert14b}, we
showed that the distribution of excess significance for the non-detections in our original survey is well behaved,
following a Gaussian with a standard deviation close to one. This suggests that errors affecting a single point (the
visibility obtained on a single baseline or on all baselines during one observation of a science target) are well
estimated by (i) our strategy to execute three consecutive observations of a science target and to include the scatter
of all 18 points (6 baselines x 3 observations) in our error estimates and (ii) the degree of partial correlation of
the data considered in our error estimates \citep{ert14b}. It also suggests that potential errors affecting the whole
observing sequence of a science target such as elevation, time dependence, or magnitude dependence are well calibrated
out by our strategy of using three to four different calibrators surrounding our science target within typically
$10^\circ$ and having very similar magnitudes to our science targets.
Situations where these effects produce a false detection can therefore be considered as very unlikely.
Unfortunately, based on this statistical argument we cannot rule out completely that such an error is present and
responsible for the measured variability of HD\,7788. We thus consider HD7788 as the first strong
candidate of significant variability, but emphasize that more data in form of denser and longer time series are needed
to confirm this result. In addition, we consider the tentative measurement of variability around HD\,210302 another
potential candidate and to characterize the variability. In both cases the u-v-coverage during all observations is
similar, so that we consider a different
u-v-coverage together with a specific excess geometry (e.g., an edge-on disk) very unlikely to be the cause of the
excess variations measured.

Although the goal of this paper is to demonstrate the persistence of the excesses, but not to discuss their nature,
we note that it has been demonstrated by \citet{mar14} that the availability of closure phase data from PIONIER
observations enables us to distinguish between the presence of a point-like companion and extended emission as a cause
for the signal.

\section{Summary and conclusions}
\label{sect_summary}

We have demonstrated that the phenomenon causing the near-infrared excess around nearby main sequence stars persists
over time scales of a few years for the majority of our detections. We have also detected with HD\,7788 the first
strong candidate of significant excess variability and with HD\,210302 another tentative candidate. In the case of
HD\,7788, the excess seems to disappear (given our sensitivity limits) within one year, but is re-detected after
another year, while in the case of HD\,210302 the excess seems to have faded away after the initial detection. We
conclude that an excess can be expected to be present around most of our targets during past follow-up observations.
Such observations to characterize detected excesses are generally not hindered by strong variability on time scales
of several months to few years. However, the potential variability in two sources demonstrates that a single star
cannot be expected to show significant excess at a given observation. Thus, we conclude that in any case a small
sample of stars needs to be observed in order to guarantee the success of a follow-up observation.

\begin{acknowledgements}
 This work has significantly benefited from the discussion at the hot dust workshop (JPL/CalTech, May~2015) organized
 by B.~Mennesson and R.~Millan-Gabet. S.~Ertel, J.-C.~Augereau, and A.~Bonsor thank the French National Research Agency
 (ANR, contract ANR-2010 BLAN-0505-01, EXOZODI) and PNP-CNES for financial support. J. Olofsson acknowledges support
 from the ALMA/Conicyt Project 31130027. PIONIER is funded by the Universit\'e Joseph
 Fourier (UJF), the Institut de Plan\'etologie et d'Astrophysique de Grenoble (IPAG), the Agence Nationale pour la
 Recherche (ANR-06-BLAN-0421 and ANR-10-BLAN-0505), and the Institut National des Science de l'Univers (INSU PNP and
 PNPS). The integrated optics beam combiner is the result of a collaboration between IPAG and CEA-LETI based on CNES
 R\&T funding. This research has made use of the Jean-Marie Mariotti Center \texttt{Aspro}\footnote{Available
 at http://www.jmmc.fr/aspro} and \texttt{SearchCal}\footnote{Available at http://www.jmmc.fr/searchcal} services, the
 latter co-developped by FIZEAU and LAOG/IPAG, and of the CDS Astronomical Databases SIMBAD and VIZIER
 \footnote{Available at http://cdsweb.u-strasbg.fr/}. The authors warmly thank everyone involved in the VLTI project.
\end{acknowledgements}

\bibliographystyle{aa}
\bibliography{bibtex}

\begin{appendix}
\section{Effects of short and variable coherence time}
\label{app_bad_obs}

The stability of PIONIER observations is generally very high, even in mediocre observing conditions. There are,
however, limits to this caused by technical limitations of the instrument. We mentioned in Sect.~\ref{sect_observations}
that in one night, \mbox{9-Aug-2013}, the observing conditions were highly variable with occasionally very large
seeing values ($>2''$) and short coherence time ($\sim$1\,ms). Since we aim at very high statistical and calibration
accuracy of our data, such conditions are problematic and we discuss the here consequences.

The fringe contrast and thus the visibility is measured with PIONIER by scanning the optical path delay (OPD) and
recording the resulting contrast over time (i.e., over OPD). This is done with very high speed (in our case, the
integration time of a single point of the scan is $\sim$1\,ms with one scan being sampled by 1024 points) in
order to freeze the effects of atmospheric turbulence. As long as the turbulence is slow enough (long enough
coherence time), this produces a very stable transfer function (TF, i.e., the contrast reached on a point source
considering all instrumental and atmospheric effects). Our experience has shown that this is the case as long
as the coherence time is longer than $\sim$2\,ms. If the coherence time drops significantly below this value,
the TF drops. This can be understood as a loss of temporal coherence of the star light due to atmospheric turbulence
that can no longer be compensated by scanning the fringes even faster because of instrumental limitations in both
scan speed and limiting magnitude.

In the night of \mbox{9-Aug-2013}, the coherence time was variable over the course of the night, ranging
from~2\,ms to below~1\,ms. For some observing sequences the coherence time was still at an acceptable level.
This means they can in principle be calibrated well. The uncertainty from this calibration as well as the
statistical uncertainty estimated from the scatter of the contrast measured on single scans are comparable to
those for data obtained in more stable conditions. However, we also need to apply a global calibration of the
night using all calibrator observations obtained in the whole night (Sect.~\ref{sect_processing}). Now, if a
fraction of these observations have been obtained with a lower TF than our science data, this will systematically
bias our data towards higher calibrated fringe contrasts. For our excess measurements, this means a systematically
lower excess measured. Since the global night calibration only introduces a correction of a few percent, the error
introduced will be only a fraction of a percent, This is however comparable to the magnitude of the signal we
intend to measure. If a significant fraction of observations were obtained during phases of short coherence time,
rejecting these data from the global calibration would result in insufficient sampling of the TF over different
pointing positions and render the whole global calibration unusable.

Since the data obtained during this night cannot be calibrated at the level of accuracy needed, and the results would
be affected by systematic errors that would bias our statistics, we discard all observations obtained during the night
of \mbox{9-Aug-2013} from our clean sample of accurate, high quality observations for the further analysis.

\end{appendix}

\end{document}